\documentclass[prl,aps,twocolumn,showpacs]{revtex4}
\usepackage{epsfig}
\usepackage{bm}

\begin{document}

\title{Nonlinear and chaotic ice ages: data vs speculations}

\author{\small  A. Bershadskii}
\affiliation{\small {ICAR, P.O.B. 31155, Jerusalem 91000, Israel}}

\begin{abstract}
It is shown that, the wavelet regression detrended fluctuations 
of the reconstructed temperature for the past 400,000 years 
(Antarctic ice cores data) are completely dominated by one-third subharmonic resonance, 
presumably related to Earth precession effect on the energy that the
intertropical regions receive from the Sun. 
Effects of Galactic turbulence on the temperature fluctuations are also discussed. 
Direct evidence of chaotic response of the atmospheric $CO_2$ dynamics to obliquity periodic 
forcing has been found in a reconstruction of atmospheric $CO_2$ data (deep sea
proxies), for the past 650,000 years. 

\end{abstract}


\maketitle

\section{Introduction}

Forcing of nonlinear systems does not always have the results that one might expect. 
Subharmonic and chaotic resonances are prominent examples of such response.  
Recent paleoclimate reconstructions provide indications of nonlinear properties of 
Earth climate at the late Pleistocene \cite{rh},\cite{sal} (the period from 0.8 Myr to present). 
Long term decrease in atmospheric $CO_2$, which could result in a change in the internal 
response of the global carbon cycle to the obliquity forcing, 
has been mentioned as one of the principal reasons for this phenomenon (see, for instance, 
\cite{berg1}-\cite{clar}). At present time one can recognize at least three problems 
of the nonlinear paleoclimate, which we will address in present paper using recent data and 
speculations. 

{\bf A}.  Reconstructed air temperature  on millennial time scales are known 
to be strongly fluctuating. See, for instance figures 1 and 2.  While the nature of the trend 
is widely discussed (in relation to the glaciation cycles) the nature of these strong fluctuations 
is still quite obscure. The problem has also a technical 
aspect: detrending is a difficult task for such strong fluctuations. 
In order to solve this problem a wavelet regression detrending method was used in present investigation. 
Then a spectral analysis of the detrended data reveals rather surprising nature of the strong 
temperature fluctuations. Namely, the detrended fluctuations of the reconstructed temperature 
are completely dominated by so-called one-third subharmonic resonance, presumably related to 
Earth precession effect on equatorial insolation. 

{\bf B}. Influence of Galactic turbulent processes on the Earth climate can be very significant for time-scales 
less than 2.5 kyr. 

{\bf C}. Nonlinearity of the Earth climate can also results in a chaotic response to an 
external periodic forcing. Already pioneering studies of the effect of external 
periodic forcing on the first Lorenz model of the chaotic climate (convection) 
revealed very interesting properties of chaotic response (see, for 
instance, \cite{golub},\cite{ahlers},\cite{fran}).  The climate, where the chaotic behavior 
was discovered for the first time, is still one of the most challenging areas for the chaotic response 
theory. One should discriminate between chaotic weather (time scales 
up to several weeks) and a more long-term climate variation. 
The weather chaotic behavior usually can be directly related to chaotic 
convection (as it was done for the first time by Lorenz), while appearance of the 
chaotic properties for the long-term climate events is a non-trivial and 
challenging phenomenon. It was suggested that such properties can play a significant role 
for glaciation cycles at multi-millennium time scales \cite{sal},\cite{hC},\cite{b}. 

Cyclic forcing, due to astronomical modulations of the solar 
input, rightfully plays a central role in the long-term climate models. 
Paradoxically, it is a very non-trivial task to find imprints of this forcing in 
the long-term climate data. It will be shown in present paper that just unusual 
properties of nonlinear and chaotic response are the main source of this problem.

\begin{figure} \vspace{-0.5cm}\centering
\epsfig{width=.45\textwidth,file=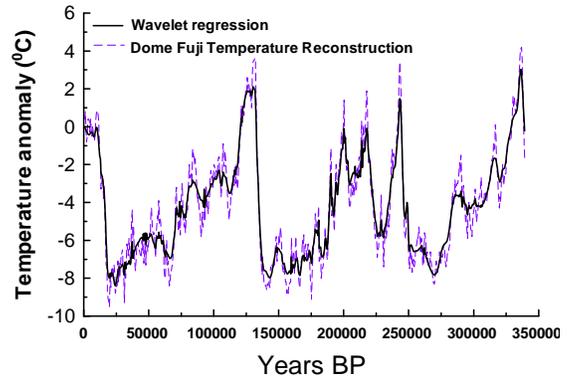} \vspace{-5cm}
\caption{The reconstructed air temperature data (dashed line) for the period 0-340 kyr. 
The data were taken from Ref. \cite{Fuji} (see also Ref. \cite{ka}). The solid curve 
(trend) corresponds to a wavelet (symmlet) regression of the data. }
\end{figure}
\begin{figure} \vspace{-1cm}\centering
\epsfig{width=.45\textwidth,file=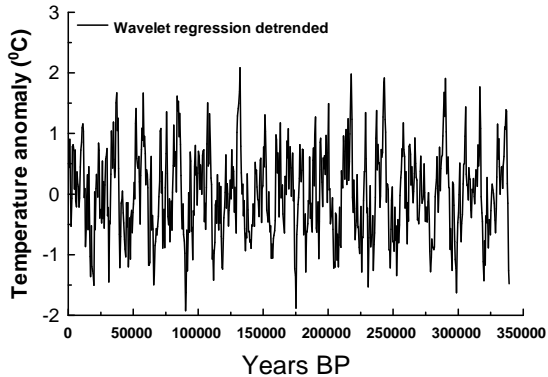} \vspace{-5cm}
\caption{The wavelet regression detrended fluctuations from the data shown in Fig. 1.}
\end{figure}

\section{Subharmonic resonance}

   Figure 1 shows reconstructed air temperature data (dashed line) for the period 0-340 kyr 
as presented at the site \cite{Fuji} (Antarctic ice cores data, see also Ref. \cite{ka}). 
The solid curve (trend) in the figure corresponds to a wavelet (symmlet) regression of the data (cf Ref. \cite{o}). Figure 2 shows corresponding detrended fluctuations, which produce 
a statistically stationary set of data. Most of the regression methods are linear in responses. 
At the nonlinear nonparametric wavelet regression one chooses a relatively small number of wavelet 
coefficients to represent the underlying regression function. A threshold method is used to keep or 
kill the wavelet coefficients. In this case, in particular, the Universal (VisuShrink) thresholding 
rule with a soft thresholding function was used. At the wavelet 
regression the demands to smoothness of the function being estimated are relaxed considerably in comparison 
to the traditional methods. 
Figure 3 shows a spectrum of the wavelet regression detrended data 
calculated using the maximum entropy method (because it provides an optimal spectral
resolution even for small data sets). One can see in this figure a small peak corresponding 
to period $\sim$ 5kyr and a huge well defined peak corresponding to period $\sim$15kyr. 
We also obtained analogous results (approximately 10\% larger) from the "Vostok" ice core data 
for period 0-420kyr (for the data description see Refs. \cite{vostok1},\cite{vostok2}). 

In order to understand underlying physics of the very characteristic picture shown in the Fig. 3 
(cf. Ref. \cite{bR}) let us imagine a forced excitable system with a large amount of loosely 
coupled degrees of freedom schematically represented by Duffing oscillators 
(which has become a classic model for analysis of 
nonlinear phenomena and can exhibit both deterministic and chaotic behavior \cite{ot}-\cite{nm} 
depending on the parameters range) with a wide range of the 
natural frequencies $\omega_0$ :

$$
\ddot{x} + \omega_0^2 x +\gamma \dot{x} +\beta x^3 = F \sin\omega t    \eqno{(1)}
$$
where $\dot{x}$ denotes the temporal derivative of $x$, $\beta$ is the strength of nonlinearity, and 
$F$ and $\omega$ are characteristic of a driving force. It is known (see for instance Ref. \cite{nm}) 
that when $\omega \approx 3\omega_0$ and $\beta \ll 1$ the equation (1) has a resonant solution 
$$
x(t) \approx a \cos\left(\frac{\omega}{3}t + \varphi \right) + \frac{F}{(\omega^2-\omega_0^2)} 
\cos \omega t   \eqno{(2)}
$$
where the amplitude $a$ and the phase $\varphi$ are certain constants. 
This is so-called one-third subharmonic resonance with the driving frequency 
$\omega$ corresponding approximately to 5kyr period (the huge peak in Fig. 3 
correspond to the first term in the right-hand side of the Eq. (2)). 
For the considered system of the oscillators an effect of synchronization can take place
and, as a consequence of this synchronization, the characteristic peaks in the spectra 
of partial oscillations coincide \cite{nl}. 
It can be useful to note, for the climate modeling, that the odd-term subharmonic resonance 
is a consequence of the reflection symmetry of the natural nonlinear oscillators 
(invariance to the transformation $x \rightarrow -x$). 

\begin{figure} \vspace{-1cm}\centering
\epsfig{width=.45\textwidth,file=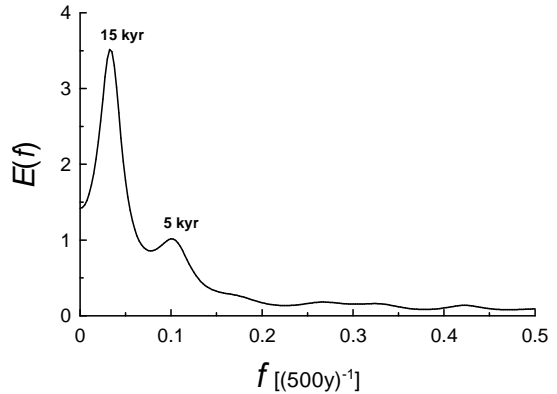} \vspace{-5cm}
\caption{Spectrum of the wavelet regression detrended fluctuations shown 
in Fig. 2.}
\end{figure}

Origin of the periodic energy input with the period $\sim$ 5kyr can be related 
to dynamics of the energy that the intertropical regions receive 
from the Sun (equatorial insolation). Indeed, it is found in Ref. \cite{blm} that 
a clear and significant 5kyr period is present in this dynamic over last 1 Myr. 
The amplitude of the 5kyr cycle in the insolation decreases rapidly when getting 
away from the equator. Using the fact that double insolation maximum and minimum arise 
in the tropical regions in the course of one year, the authors of the Ref. \cite{blm} 
speculated that this period in seasonal amplitude of equatorial insolation is determined 
by fourth harmonic of the Earth precession cycle. It should be noted, that the idea of a significant 
role of tropics in generating long-term climatic variations is rather a new one (see Ref. \cite{blm} 
for relevant references). In Ref. \cite{hag}, for instance, the authors speculated that the 
high frequency climate variability (in the millennial time scales) could be related to high 
sensitivity of the tropics to summer time insolation. Then, the oceanic advective transport 
could transmit an amplified response of tropical precipitation and temperature to high latitudes. 
Physical mechanism of this amplification is still not clear and the above discussed one-third subharmonic 
resonance can be a plausible possibility (in this respect it is significant, that we used the 
Antarctic data).  

\section{Galactic turbulence and the temperature fluctuations}

Since the high frequency part of the spectrum is corrupted by strong fluctuations (the Nyquist frequency 
equals 0.5 [$(500y)^{-1}$]), it is interesting to look at corresponding autocorrelation function $C(\tau)$ 
in order to understand what happens on the millennial time scales.
Figure 4 shows a relatively small-times part of the correlation function defect. 
The ln-ln scales have been used in this figure in order to show a power law (the straight line) 
for structure function: $\langle (v(t+\tau)- v(t))^2 \rangle$ :
$$
1- C(\tau ) \propto  \langle (v(t+\tau)- v(t))^2 \rangle \propto \tau^{2/3}  \eqno{(3)}
$$  
This power law: '2/3', for structure function (by virtue of the Taylor hypothesis 
transforming the time scaling into the space one \cite{my},\cite{b1}) 
is known for fully developed turbulence as Kolmogorov's power law. 

\begin{figure} \vspace{-0.5cm}\centering
\epsfig{width=.45\textwidth,file=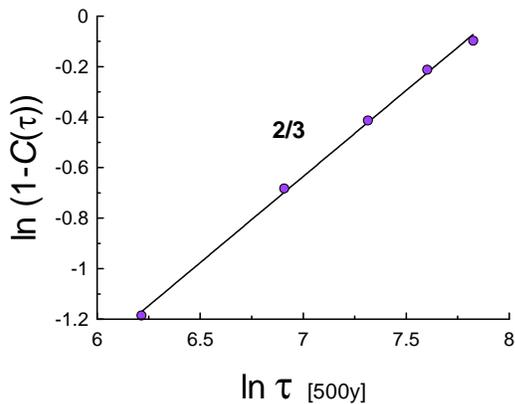} \vspace{-4.5cm}
\caption{A small-time-scales part of the autocorrelation function defect of the wavelet regression 
detrended fluctuations from the data shown in Fig. 2. The straight 
line indicates the Kolmogorov's '2/3' power law for the structure function in ln-ln scales.}
\end{figure}
Although, the scaling interval is short, the value of the exponent is rather intriguing. 
This exponent is well known in the theory of fluid (plasma) turbulence and corresponds to so-called 
Kolmogorov's cascade process. This process is very universal for turbulent fluids 
and plasmas \cite{gibson},\cite{clv}. 
For turbulent processes on Earth and in Heliosphere 
the Kolmogorov-like scaling with such large time scales certainly cannot exist. Therefore, one should think about 
a Galactic origin of Kolmogorov turbulence with such large time scales (let us recall that diameter of the Galaxy 
is approximately 100,000 light years). 
This is not surprising if we recall possible role of the 
galactic cosmic rays for Earth climate (see, for instance, \cite{b},\cite{uk}-\cite{kirk}). 
In this respect, it should be also noted that the '2/3' scaling law can reflect not the velocity 
field of the galactic interstellar media but the turbulent electron density field 
(presumably produced by supernova) according to the Obukhov-Corrsin turbulent mixing process (see, for instance, 
Ref. \cite{gibson1},\cite{gs} and references therein).

\section{Atmospheric $CO_2$ dynamics at multi-millennium time scales}

The angle between Earth's rotational axis and the normal to the plane of its orbit 
(known as {\it obliquity}) varies periodically between 22.1 degrees and 24.5 degrees 
on about 41,000-year cycle. Such  multi-millennium timescale changes in orientation change 
the amount of solar radiation reaching the Earth in different latitudes. 
In high latitudes the annual mean insolation (incident solar radiation) decreases with 
obliquity, while it increases in lower latitudes. Obliquity forcing effect is maximum 
at the poles and comparatively small in the tropics. 
Milankovi\'{c} theory suggests that lower obliquity, leading to reduction in summer insolation 
and in the mean insolation in high latitudes, favors gradual accumulation of ice and snow
leading to formation of an ice sheet. The obliquity forcing on Earth 
climate is considered as the primary driving force for the cycles of glaciation (see for a recent 
review \cite{rh}). Observations show that glacial changes from -1.5 to -2.5 Myr (early Pleistocene) 
were dominated by 
41 kyr cycle \cite{hC},\cite{rm},\cite{h1}, whereas the period from 0.8 Myr to present (late Pleistocene)
is characterized by approximately 100 kyr glacial cycles \cite{hays},\cite{im}. 
While the 41 kyr cycle of early Pleistocene glaciation is readily related to the 41 kyr period 
of Earth's obliquity variations 
the 100 kyr period of the glacial cycles in late Pleistocene still presents a serious problem. 
Influence of the obliquity variations on global climate started amplifying around 2.5 Myr,    
and became nonlinear at the late Pleistocene. Long term decrease in atmospheric $CO_2$, which 
could result in a change in the internal response of the global carbon cycle to the obliquity forcing, 
has been mentioned as one of the principal reasons for this phenomenon (see, for instance, 
\cite{berg1}-\cite{clar}). Therefore, investigation of the historic variability in 
atmospheric $CO_2$ can be crucial for understanding the global climate changes at millennial 
timescales. 
\begin{figure} \vspace{-0.5cm}\centering
\epsfig{width=.45\textwidth,file=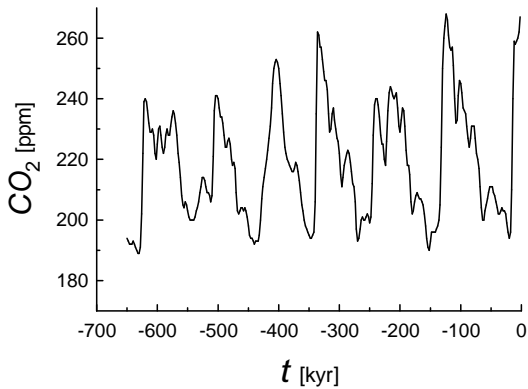} \vspace{-4.5cm}
\caption{A reconstruction of atmospheric $CO_2$ based on deep-sea proxies, 
for the past 650kyr. The data were taken from \cite{berg-data}.}
\end{figure}

\begin{figure} \vspace{-0.5cm}\centering
\epsfig{width=.45\textwidth,file=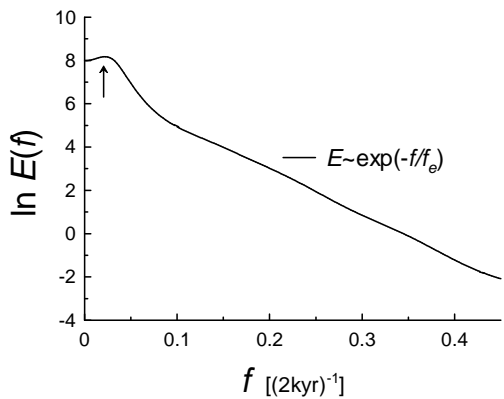} \vspace{-4.5cm}
\caption{Spectrum of atmospheric $CO_2$ fluctuations for the data shown in Fig. 5}
\end{figure}
 Figure 5 shows a reconstruction of atmospheric $CO_2$ based on deep-sea proxies, 
for the past 650kyr (the data taken from \cite{berg-data}). Resolution of the data set is 
2kyr. Fluctuations with time-scales less than 2kyr could be rather large 
(statistically up to 308ppm \cite{berg-data}), but they are smoothed by the resolution. 
Figure 6 shows a power spectrum of the data set calculated using the maximum entropy method. 
The spectrum exhibits a peak indicating a periodic component 
(the arrow in the Fig. 6 indicates a 100kyr period) and a broad-band part with exponential decay
$$
E(f) \sim e^{-f/f_e}   \eqno{(4)}
$$
A semi-logarithmic
plot was used in Fig. 6 in order to
show the exponential decay more clearly (at this plot the
exponential decay corresponds to a straight line). Both
stochastic and deterministic processes can result in the
broad-band part of the spectrum, but the decay in the
spectral power is different for the two cases. The exponential
decay indicates that the broad-band spectrum
for these data arises from a deterministic rather than a
stochastic process. For a wide class of deterministic
systems a broad-band spectrum with exponential
decay is a generic feature of their chaotic solutions 
Refs. \cite{oht}-\cite{fm}. Let us recall that the discussed above Duffing oscillators 
can exhibit both deterministic and chaotic behavior \cite{ot}-\cite{nm},\cite{oht} depending on the 
parameters range.

Nature of the exponential decay of the power spectra
of the chaotic systems is still an unsolved mathematical
problem. A progress in solution of this problem
has been achieved by the use of the analytical continuation
of the equations in the complex domain (see, for 
instance, \cite{fm}). In this approach the exponential decay
of chaotic spectrum is related to a singularity in the
plane of complex time, which lies nearest to the real axis.
Distance between this singularity and the real axis determines
the rate of the exponential decay. For many interesting cases 
chaotic solutions are analytic in a finite strip around the real time axis. 
This takes place, for instance for attractors bounded in the real 
domain (the Lorentz attractor, for instance). 
In this case the radius of convergence of the Taylor series 
is also bounded (uniformly) at any real time. 
If parameters of the dynamical system fluctuate periodically 
around their mean values with period $T_e$, then the restriction of the Taylor series 
convergence (at certain conditions) is determined by the period of the 
parametric modulation, and the width of the analytic strip around 
real time axis equals $T_e/2\pi$ \cite{b}. Let us consider, for 
simplicity, solution $u(t)$ with simple poles only, and to define the Fourier 
transform as follows
$$
u(\omega) =(2\pi)^{-1/2} \int_{-T_e/2}^{T_e/2} dt~e^{-i \omega t} u(t)  \eqno{(5)}
$$  
Then using the theorem of residues
$$
u(\omega) =i (2\pi)^{1/2} \sum_j R_j \exp (i \omega x_j -|\omega y_j|)  \eqno{(6)}
$$
where $R_j$ are the poles residue and $x_j + iy_j$ are their location in the relevant half
plane, one obtains asymptotic behavior of the spectrum $E(\omega)= |u(\omega)|^2$ at large $\omega$
$$
E(\omega) \sim \exp (-2|\omega y_{min}|)  \eqno{(7)}
$$
where $y_{min}$ is the imaginary part of the location of
the pole which lies nearest to the real axis. Therefore, exponential
decay rate of the broad-band part of the system
spectrum, Eq. (4), equals the period of the parametric forcing.

The chaotic spectrum provides two different characteristic
time-scales for the system: a period corresponding to
fundamental frequency of the system, $T_{fun}$, and a period
corresponding to the exponential decay rate, $T_e = 1/f_e$
(cf. Eq. (4)). The fundamental period $T_{fun}$ can be estimated
using position of the low-frequency peak, while the
exponential decay rate period $T_e = 1/f_e$ can be estimated
using the slope of the straight line of the broad-band part
of the spectrum in the semi-logarithmic representation (Fig. 6). 
From Fig. 6 we obtain 
$T_{fun} \simeq 95 \pm 8$ kyr (the peak is quite broad 
due to small data set) and $T_e \simeq 41 \pm 1$ kyr 
(the estimated errors are statistical ones).

Thus, the obliquity period of 41 kyr is still a dominating factor in the chaotic $CO_2$ 
fluctuations, although it is hidden for linear interpretation of the power spectrum. 
In the nonlinear interpretation the additional period $T_{fun}\simeq 100$ kyr may correspond 
to the fundamental frequency of the underlying nonlinear dynamical system and it determines the 
apparent 100 kyr 'periodicity' of the glaciation cycles for the last 650 kyr 
(cf Refs. \cite{sal},\cite{ru},\cite{hug1} and references therein). \\

The data were provided by World Data Center for Paleoclimatology, Boulder and 
NOAA Paleoclimatology Program. I also acknowledge that a software provided 
by K. Yoshioka was used at the computations.


\begin{thebibliography}{99}
\bibitem{rh} M. Raymo and P. Huybers, Unlocking the mysteries of the ice ages, 
Nature {\bf 451}, 284-285 (2008)
\bibitem{sal} B. Saltzman, Dynamical paleoclimatology : generalized theory of global
climate change. (Academic Press, San Diego, 2001).
\bibitem{berg1} A. Berger, X. Li, and M.F. Loutre, Modelling northern hemisphere ice volume over
the last 3 Ma. Quaternary Science Reviews, {\bf 18}, 1-11 (1999).
\bibitem{ru} W.F. Ruddiman, Orbital insolation, ice volume, and greenhouse gases.
Quaternary Science Reviews, {\bf 22}, 1597-1629 (2003).
\bibitem{clar} P. Clark, D. Archer, D. Pollard, J. Blum, J., et al., 
The middle Pleistocene transition: characteristics, mechanisms, and implications for longterm
changes in atmospheric $CO_2$, Quaternary Sci. Rev., {\bf 25}, 3150-3184 (2006).
\bibitem{golub} J.P. Gollub, and S. V. Benson, 
Chaotic response to periodic perturbations of a convecting flow, Phys. Rev. Lett., 
{\bf 41}, 948-951 (1978). 
\bibitem{ahlers} G. Ahlers, P. C. Hohenberg, and M. L\"{u}cke,  
Thermal convection under external modulation of the driving force. 
I. The Lorenz model. Phys. Rev. A, {\bf 32}, 3493-3518. (1985). 
\bibitem{fran} M. Franz, and M. Zhang, 
Suppression and creation of chaos in a periodically forced Lorenz system. 
Phys. Rev. E, {\bf 52}, 3558-3565 (1995). 
\bibitem{hC} P. Huybers, Early Pleistocene glacial cycles and the integrated summer insolation forcing,
Science, {\bf 313}, 508-511 (2006).
\bibitem{b} A. Bershadskii, Chaotic climate response to long-term solar forcing variability, EPL
(Europhys. Lett.), {\bf 88}, 60004 (2009).
\bibitem{Fuji} The data are available at http://www.ncdc.noaa.gov/paleo/metadata/noaa-icecore-6076.html
\bibitem{ka} K. Kawamura et al., Northern Hemisphere forcing of climatic cycles in Antarctica over
the past 360,000 years,  Nature, {\bf 448}, 912-916 (2007).
\bibitem{o} T. Ogden, Essential Wavelets for Statistical Applications and Data Analysis 
(Birkhauser, Basel, 1997).
\bibitem{vostok1} J.R. Petit, J.R., et al., Vostok Ice Core Data for 420,000 Years, IGBP PAGES/World Data Center 
for Paleoclimatology Data Contribution Series \#2001-076.  NOAA/NGDC Paleoclimatology Program, Boulder CO, USA (2001). 
\bibitem{vostok2} J. Jouzel J., et al., Climate and Atmospheric History of the Past 420,000 years from the 
Vostok Ice Core, Antarctica, Nature, {\bf 399}, 429-436 (1999).
\bibitem{bR} A. Bershadskii, Subharmonic resonance of global climate to solar forcing, arXiv:1002.1024 (2010).
\bibitem{ot} E. Ott, Chaos in Dynamical Systems (Cambridge University Press, 2002).
\bibitem{ph} D. Permann and I. Hamilton, Wavelet analysis of time series for the Duffing oscillator: 
The detection of order within chaos, Phys. Rev. Lett., {\bf 69}, 2607 (1992).
\bibitem{bh} V. Brunsden and P. Holmes, Power spectra of strange attractors near homoclinic orbits, 
Phys. Rev. Lett., {\bf 58}, 1699 (1987). 
\bibitem{nm} A.H. Nayfeh and D.T. Mook, "Nonlinear Oscillations" (John Wiley \& Sons, 
a Wiley-Interscience Publication, 1979).
\bibitem{nl} Yu.I. Neimark and P.S. Landa, Stochastic and Chaotic Oscillations,
(Dordrecht, Kluwer, 1992).
\bibitem{blm} A. Berger, M.F. Loutre, and J. L. Melice, Equatorial insolation: from precession
harmonics to eccentricity frequencies, Clim. Past Discuss., {\bf 2}, 519-533 (2006).
\bibitem{hag} T.K. Hagelberg, G. Bond, and P. de Menocal, Milankovitch band forcing of sub-Milankovitch
climate variability during the Pleistocene, Paleoceanography, {\bf 9}, 545-558 (1994).
\bibitem{my} A. S. Monin and A. M. Yaglom, Statistical Fluid Mechanics, Vol. II (MIT
Press, Cambridge, 1975).
\bibitem{b1} A. Bershadskii, Multiscaling of Galactic Cosmic Ray Flux, 
Phys. Rev. Lett., {\bf 90}, 041101 (2003) (see also arXiv:astro-ph/0305453).
\bibitem{gibson} C.H Gibson, Kolmogorov Similarity Hypotheses for Scalar Fields:
Sampling Intermittent Turbulent Mixing in the Ocean and Galaxy, Proc. Roy. Soc. Lond. {\bf 434}, 149 (1991).
\bibitem{clv} J. Cho, A. Lazarian, and E.T. Vishniac, Simulations of MHD Turbulence in a Strongly Magnetized 
Medium, Astrophys. J. {\bf 564}, 291-301  (2002) (see also arXiv:astro-ph/0205286).
\bibitem{uk} I. G. Usoskin and G. A. Kovaltsov, Cosmic Ray Induced
Ionization in the Atmosphere: Full Modeling and
Practical Applications, J. Geophys. Res., {\bf 111}, D21206, (2006).
\bibitem{shav} N.J. Shaviv, On climate response to changes in the cosmic ray flux and radiative budget, 
J. Geophys. Res. {\bf 110}, A08105 (2005). 
\bibitem{kirk} J. Kirkby, Cosmic Rays and Climate, Surveys in Geophysics, {\bf 28}, 333-375 (2007) (see also  arXiv:0804.1938). 
\bibitem{gibson1} C.H. Gibson, R.N. Keeler, V.G. Bondur, et al., Submerged turbulence detection with optical satellites, Proc. of SPIE, {\bf 6680}, 6680-33 (2007) (see also arXiv:0709.0074v2 [astro-ph]).
\bibitem{gs} C.H. Gibson and R.E. Schild, Hydro-Gravitational-Dynamics of Planets and Dark Energy, 
J. Appl. Fluid. Mech., {\bf 2} 35-41 (2009).
\bibitem{rm} M. Raymo, and K. Nisancioglu, The 41 kyr world: Milankovitch's other unsolved mystery, 
Paleoceanography, {\bf 18}, 1011, (2003) 
\bibitem{h1} P. Huybers, Glacial variability over the last two million years: an extended depth-derived age-
model, continuous obliquity pacing, and the Pleistocene progression, Quaternary Sci. Rev.,
{\bf 26}, 37-55 (2007).
\bibitem{hays} J. Hays, J. Imbrie, and N. Shackleton, Variations in the Earth's Orbit: pacemaker of the Ice
Ages, Science, {\bf 194}, 1121-1132 (1976).
\bibitem{im} J. Imbrie, E.A. Boyle, S.C. Clemens, et al., On the structure and 
origin of major glaciation cycles. 1. Linear responses to Milankovitch forcing, Paleoceanography,
{\bf 7}, 701-738 (1992).
\bibitem{berg-data} W.H. Berger, Database for reconstruction of atmospheric CO2 in the
Milankovitch Chron, IGBP PAGES/World Data Center-A for Paleoclimatology
Data Contribution Series \# 96-031.  NOAA/NGDC Paleoclimatology Program,
Boulder CO, USA (see also W.H.Berger, T.Bickert, M.K.Yasuda, G.Wefer, Reconstruction of
atmospheric $CO_2$ from ice-core data and the deep-sea record of Ontong Java
plateau: the Milankovitch chron. Geologische Rundschau, {\bf 85}, 466-495 (1996)).
\bibitem{oht} N. Ohtomo, K. Tokiwano, Y. Tanaka, A. Sumi, S. Terachi, and H. Konno, 
Exponential Characteristics of Power Spectral Densities Caused by Chaotic Phenomena, 
J. Phys. Soc. Jpn. {\bf 64} 1104-1113  (1995). 
\bibitem{fa} J. D. Farmer, Chaotic attractors of an infinite dimensional dynamic system, Physica D, {\bf 4}, 
366-393 (1982).
\bibitem{sig} D.E. Sigeti, Survival of deterministic dynamics in the presence of noise and the
exponential decay of power spectrum at high frequencies. Phys. Rev. E, {\bf 52}, 2443-2457 (1995).
\bibitem{hav} L. A. Safonov, E. Tomer, V. V. Strygin, Y. Ashkenazy, and S. Havlin, 
Delay-induced chaos with multifractal attractor in a traffic flow model, Europhys. Lett., {\bf 57}, 
151-157 (2002).
\bibitem{fu} F. Fucito, F. Marchesoni, E. Marianari, G. Parisi,
L.Peliti, S. Ruffo, and V. Vulpiani, Approach to equilibrium
in a chain of nonlinear oscillators, J. Physique,
{\bf 43}, 707-713 (1982).
\bibitem{fm} U. Frisch and R. Morf, Intermittency in non-linear dynamics
and singularities at complex times, Phys. Rev.
{\bf 23}, 2673-2704 (1981).
\bibitem{hug1} P. Huybers,  Pleistocene glacial variability as a chaotic
response to obliquity forcing, Clim. Past Discuss., {\bf 5}, 237-250 (2009).
\end{thebibliography}
\end{document}